\documentclass[runningheads]{llncs}

\usepackage[english]{babel}
\usepackage[T1]{fontenc}
\usepackage{todonotes}
\usepackage{tikz}
\usepackage{listings}
\usepackage{hyperref}
\usepackage{cleveref}
\AtBeginDocument{}
\usepackage{xcolor}
\definecolor{mGreen}{HTML}{8fbcbb}
\definecolor{mDarkGray}{HTML}{4c566a}
\definecolor{mLightBg}{HTML}{E6EFEF}
\definecolor{mGray}{rgb}{0.5,0.5,0.5}
\definecolor{mPurple}{rgb}{0.58,0,0.82}
\definecolor{mBlue}{HTML}{5e81ac}
\definecolor{mOrange}{HTML}{d08770}
\definecolor{mYellow}{HTML}{ebcb8b}
\definecolor{mRed}{HTML}{bf616a}

\lstdefinestyle{mcrl2}{
  language=,
  basicstyle=\ttfamily\footnotesize\lst@ifdisplaystyle\scriptsize\fi,
  keywords={sort,act,proc,init,sum,allow,hide,comm,block,rename,var,eqn,map,struct,cons,forall,exists},
  morekeywords={Set,Nat,Bool,List,true,false,whr,end,delta,tau},
  keywordstyle=\bfseries\color{mBlue},
  stringstyle=\color{mDarkGray},
  commentstyle=\color{mOrange},
  sensitive=true,
  morecomment=[l]{\%},
  captionpos=b,
  commentstyle=\itshape,
  stringstyle=\ttfamily,
  showstringspaces=false,
  numbers=none,
  numberstyle=\tiny,
  breaklines=true,
  frame=single,
  mathescape=true
}

\usepackage{svg}

\title{A Complete Formal Specification and Verification of the BESW software control system of the Maeslant Storm Surge Barrier\thanks{~\vspace{-4.5ex}\\
\begin{minipage}[t]{0.45\linewidth}Research is supported by the Interreg North Sea project STORM\_SAFE\\
\scriptsize\url{www.interregnorthsea.eu/stormsafe}
\end{minipage}\hfill
\begin{minipage}[c]{0.5\linewidth}
\vspace{3.5ex}~\\\includegraphics[width=\linewidth]{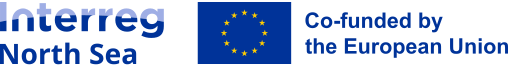}\end{minipage}}}
\author{Adrian Beers\orcidID{0009-0001-2955-3659} \and Jore Booy\inst{1}\orcidID{0009-0006-2233-6607} \and Jan Friso Groote\inst{1}\orcidID{0000-0003-2196-6587} \and Johan van den Bogaard\inst{2}\orcidID{0009-0005-9018-250X} \and Mark Bouwman\inst{2}\orcidID{0000-0002-5131-008X}
}

\titlerunning{Formal Specification and Verification of the Maeslant Barrier}

\authorrunning{Beers et al.}

\institute{$^1$Eindhoven University of Technology, Eindhoven, The Netherlands\\$^2$Rijkswaterstaat, The Netherlands}
\begin{document}

\maketitle

\begin{abstract}
The Maeslant Barrier is a storm surge barrier that protects Rotterdam and its harbour from storm surges in
  the North Sea. Its software control consists of three major components, one of which is BesW. BesW is responsible for all the movements of the barrier except for pushing and pulling it. In this document, we report on the complete
  formal specification of BesW in mCRL2. All its behaviour has been specified, including manual and testing modes. Furthermore, all fault situations have been taken into account. The formalisation allows formal verification of all behavioural properties, formulated in the modal $\mu$-calculus, with the constraints that water levels only have a restricted number of values and not all combinations of failures of pumps and valves are allowed. 
  \keywords{Formal modelling. Model checking. mCRL2. Maeslant Barrier. BesW}
\end{abstract}

%\begin{figure}[b!]
%    \centering
%    \includegraphics[width=0.5\linewidth]{logo_interreg.png}
%\end{figure}

\section{Introduction}

The Maeslant Barrier is a storm surge barrier in the Netherlands. It is one of the largest movable structures on earth. It is located at the end of the rivers Maas and Rhine in a human-made canal called the Nieuwe Waterweg,  connecting the North Sea with the largest harbour in Europe, namely Rotterdam. In case the sea level rises above a threshold, two huge retaining walls are pushed into the Nieuwe Waterweg and lowered onto a concrete base to prevent water from flowing inland.

The software control of the barrier is up for replacement. In order to do this,
Rijkswaterstaat, the Dutch governmental body responsible for flood protection,
desires to have complete and correct formal descriptions of the prospective
software control. 

In this paper, we embark upon this request and ask ourselves whether these control system can be fully modelled, preferably in such a way that correctness properties can be checked. We use the formal specification language and tool mCRL2 for this purpose due to its versatility and its successful use in many comparable large-scale modelling projects \cite{grooteModelingAnalysisCommunicating2014a}. The goal of this case study is to determine to what extent formal modelling and verification can be applied to the software controlling the Maeslant Barrier. 

More concretely, we looked into BesW (BesturingsSysteem Nieuwe Waterweg). The software control of the Maeslant Barrier has three main parts. The first part, BOS (Beslis en Ondersteunend Systeem), is the overarching software part responsible for the decisions to close and open the barrier. It instructs the second part, BesW, in terms of the major phases of a closure. BesW is responsible for controlling pumps, valves, doors, locks, a.o., to let the barrier float and sink, flood the dry dock, and open and close the dry docks' door. BesW instructs the third main part, BesL, which is the control system for the locomobile, to move the wall into or out of the river. There are two symmetrical instances for both BesW and BesL for the northern and southern retaining walls. 

Our results demonstrate that a complete formal model and property verification of the BESW system is feasible and effective. Not only could the full behaviour of the software control system be modelled, including failing hardware, and rest and testing modes (ITO), but it was also possible to verify or disprove all properties that were formulated on this behaviour. In the model, status and log messages were omitted. In the verification the water levels were restricted to a representative subset and not all combinations of failing pumps and valves were taken into account. 

Unfortunately, for reasons of national security, the precise formal model of the control of the Maeslant Barrier is declared confidential by Rijkswaterstaat. A partial public description of the Maeslant Barrier and the verified properties can be found in \cite{beersSpecificationVerificationMain2024} on which this article is based. We necessarily have to limit ourselves to describing how the model is constructed and which kind of properties have been verified. But we still believe that reporting on this model is very important as due to its complexity and importance, it can be seen as a landmark in the field of formal methods.  

\section{Related work}

Dutch water defence works have been the subject of study by formal methods. The decision making system (BOS) of the Maeslant Barrier has been verified using Promela for the communication protocols and Z~\cite{spivey1992z} for system behaviour~\cite{tretmansSoftwareEngineeringFormal2001}. Based on the Z~specification, an informal C++ implementation was created. Some years later, Madlener et al.~\cite{madlenerFormalVerificationStudy2010} studied the code for a specific component, `Determine Excessive Water Level', and compared the code to the Z~specification by translating the C++ code to the PVS theorem prover. Some discrepancies were found between the C++ code and the Z~specification.

The control system of the locomobile component of the Maeslant Barrier has been modelled in mCRL2~\cite{visscherFormalModellingVerification2023}. In this work, some properties could not be checked as the size of the state space proved to be an issue for verification. Most of the remaining properties were valid, and four properties were found to be false due to errors in the functional description of the system. Sewberath-Misser~\cite{sewberath-misserImprovingReliabilityMaeslant2022} analysed and improved the mechanical failure probability of a closure based on various storm scenarios.
Van der Meulen and Clement~\cite{vandermeulenFormalMethodsSpecification1999} verified the decision procedure for closing the barriers of the Eastern Scheldt Barrier (Oosterscheldekering) using HOL90. 
Wilschut~\cite{wilschutSystemSpecificationDesign2018} verified safety and liveness properties of the Prinses Ma\-rij\-ke Locks in ESL~\cite{wilschutBridgingGapRequirements2024}, a specification language for system component verification.
Reijnen et al.~\cite{reijnenSynthesisImplementationSupervisory2019} generated PLC code for the Oisterwijksebaan bridge based on a supervisor model. Short of some minor complications due to the interfaces of the supervisor and the sensors, all tests performed were successful. 

\section{Maeslant Barrier}

The Maeslant Barrier is the largest automated storm surge barrier in the Netherlands.
The barrier is open during normal operation to allow for maritime traffic and closes automatically when the water level passes a predetermined threshold.

\begin{figure}[htb]
    \centering
  \includegraphics[width=0.7\textwidth]{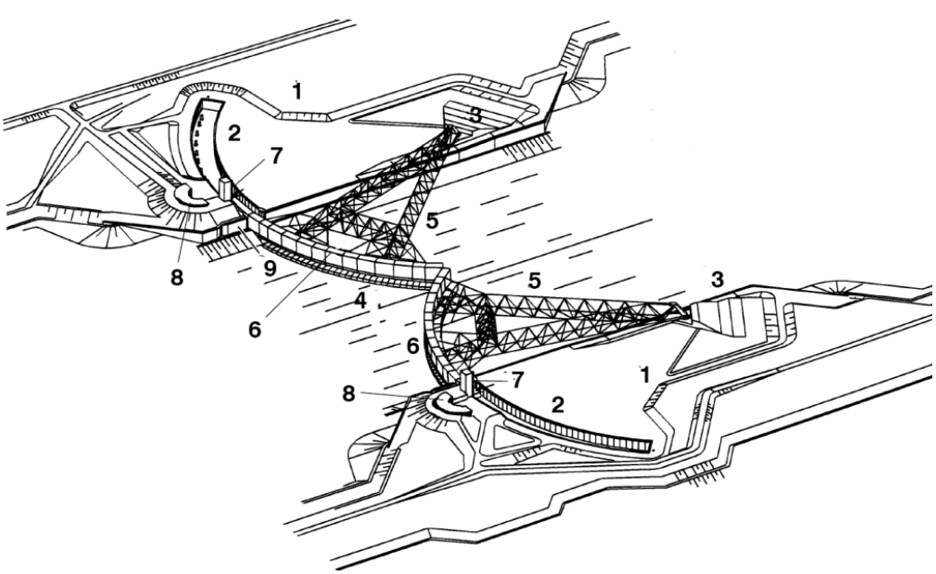}
  \caption{The Maeslant Barrier}
  \label{fig:barrier}
\end{figure}

The barrier consists of two symmetric sides, referred to as North and South. They both contain a sector door, which consists of a retaining wall (6), a ball joint (3) and trusses (5) connecting them. Furthermore, there is a parking dock (2) and a locomobile (7). Closing the barrier consists roughly of flooding the docks and opening the docks' doors, after which the locomobiles push the retaining walls into the river, where they sink to the bottom. Opening the barrier follows the reverse sequence with the locomobiles pulling the walls.

\begin{figure}[htbp]
    \centering
    \includegraphics[width=0.8\linewidth]{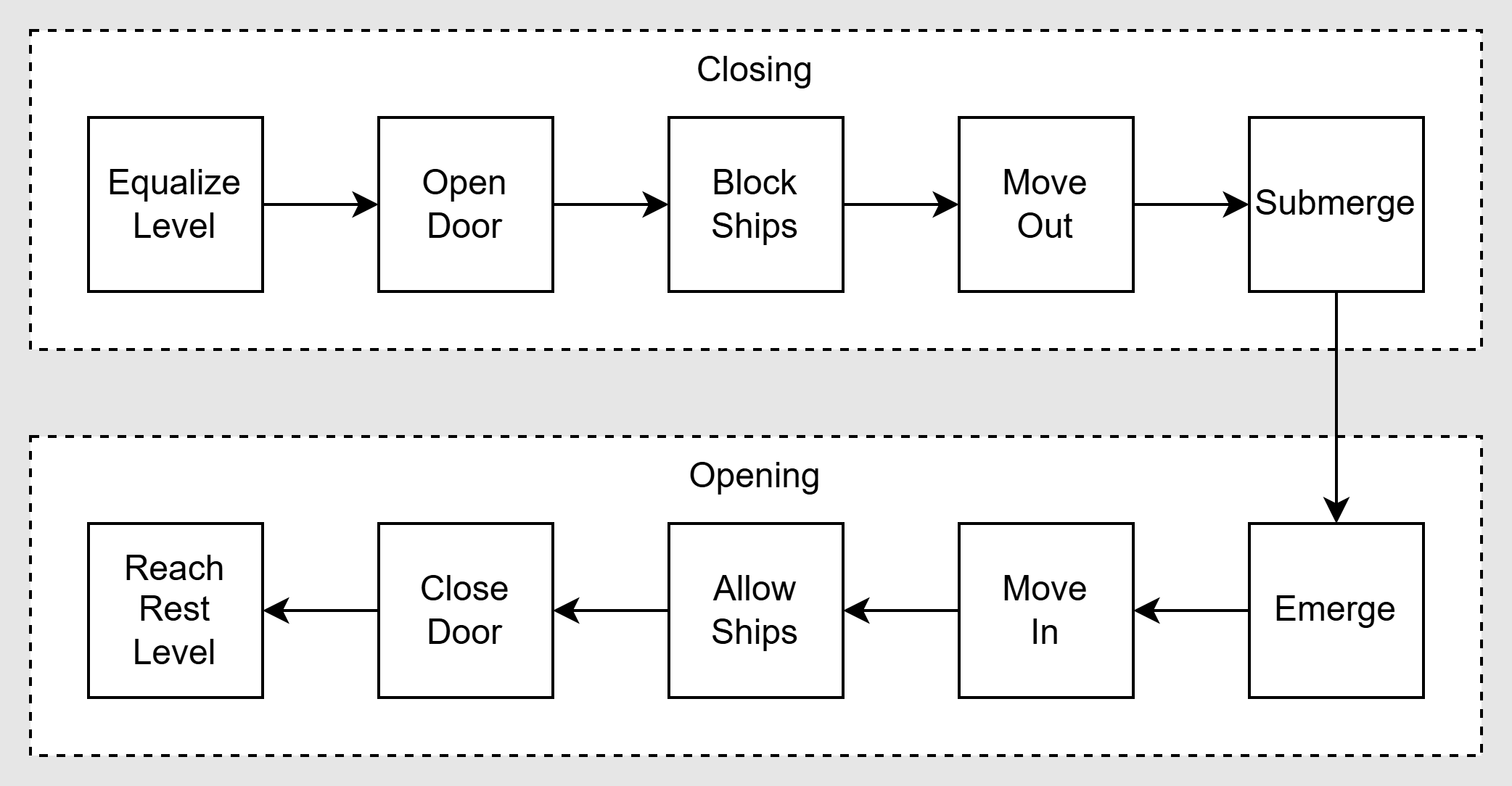}
    \caption{The operational processes}
    \label{fig:operational-processes}
\end{figure}

The system can be set to the regular operational mode, a mode for waiting for instructions (Rest) or a testing mode (ITO). 
The operational processes that describe the behavioural steps the system takes to open and close the barrier can be seen in~Fig.~\ref{fig:operational-processes}. Most are self-explanatory. `Equalise Level' and `Reach Rest Level' are concerned with the difference in water level of the river and the dock. 

\begin{figure}[hbtp]
    \centering
    \includegraphics[width=0.7\linewidth]{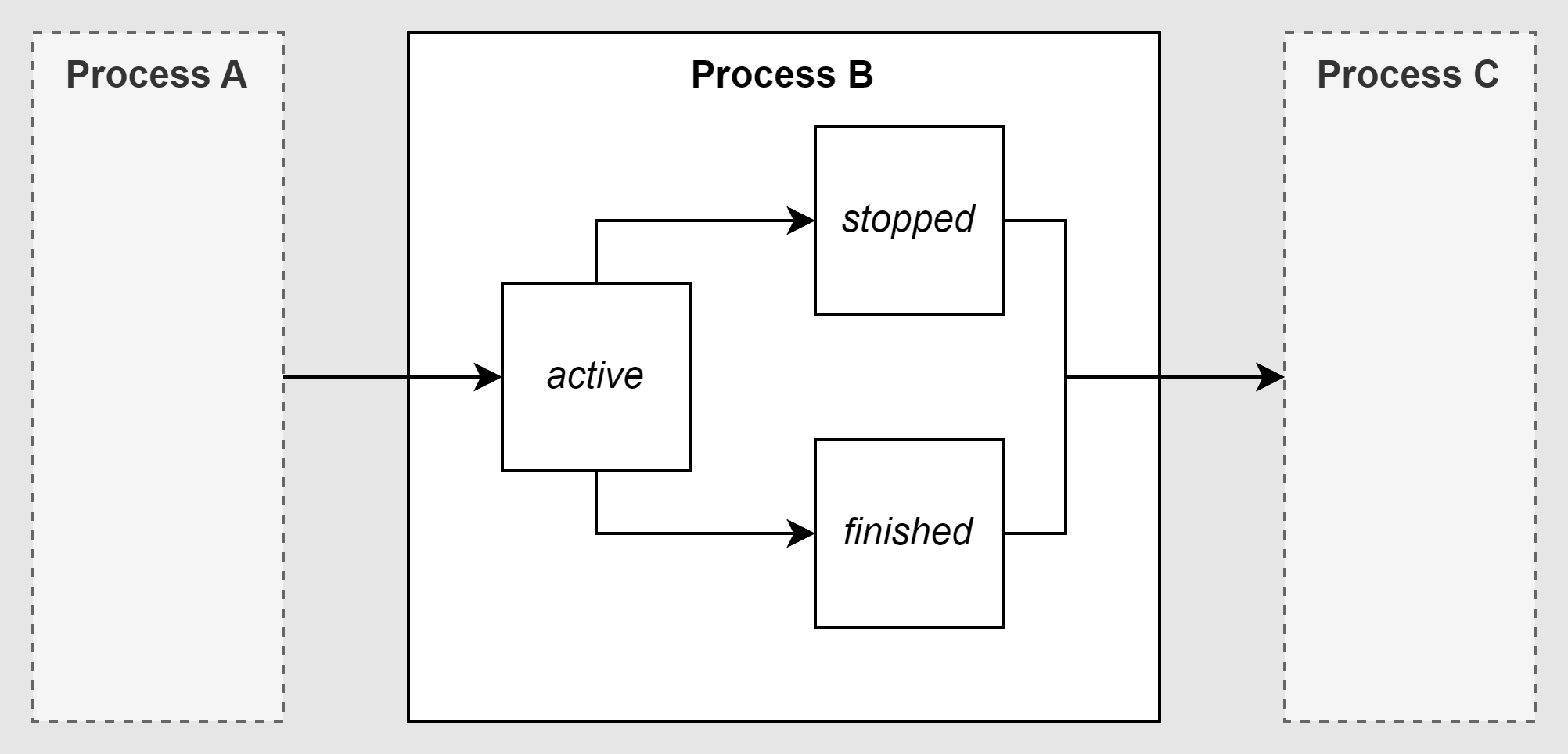}
    \caption{Process modes}
    \label{fig:process-mode}
\end{figure}

Processes are activities of an individual system, which may take some time to finish. The current state of a process is the current process `mode' (see Fig.~\ref{fig:process-mode}). A process first starts in `active' mode. The process transitions to the `finished' state when a finishing condition is reached. The process transitions to the `stopped' state when a stop command is received. BesW can start a new process when the current processes is either `stopped' or `finished'.

\subsection{BesW}
BesW (Besturingssysteem Nieuwe Waterweg) is the main control system of the Maeslant Barrier. The system controls the retaining walls, the ball joints, the parking docks and the locomobiles (BesL). BesW for instance instructs BesL with respect to movement, the spring setting and ventilator settings. There is one instance of BesW for each side of the barrier, BesW North and BesW South. Both instances of BesW communicate with each other to exchange statuses and measurements.

\begin{figure}[htbp]
  \centering
  \includegraphics[width=0.5\textwidth]{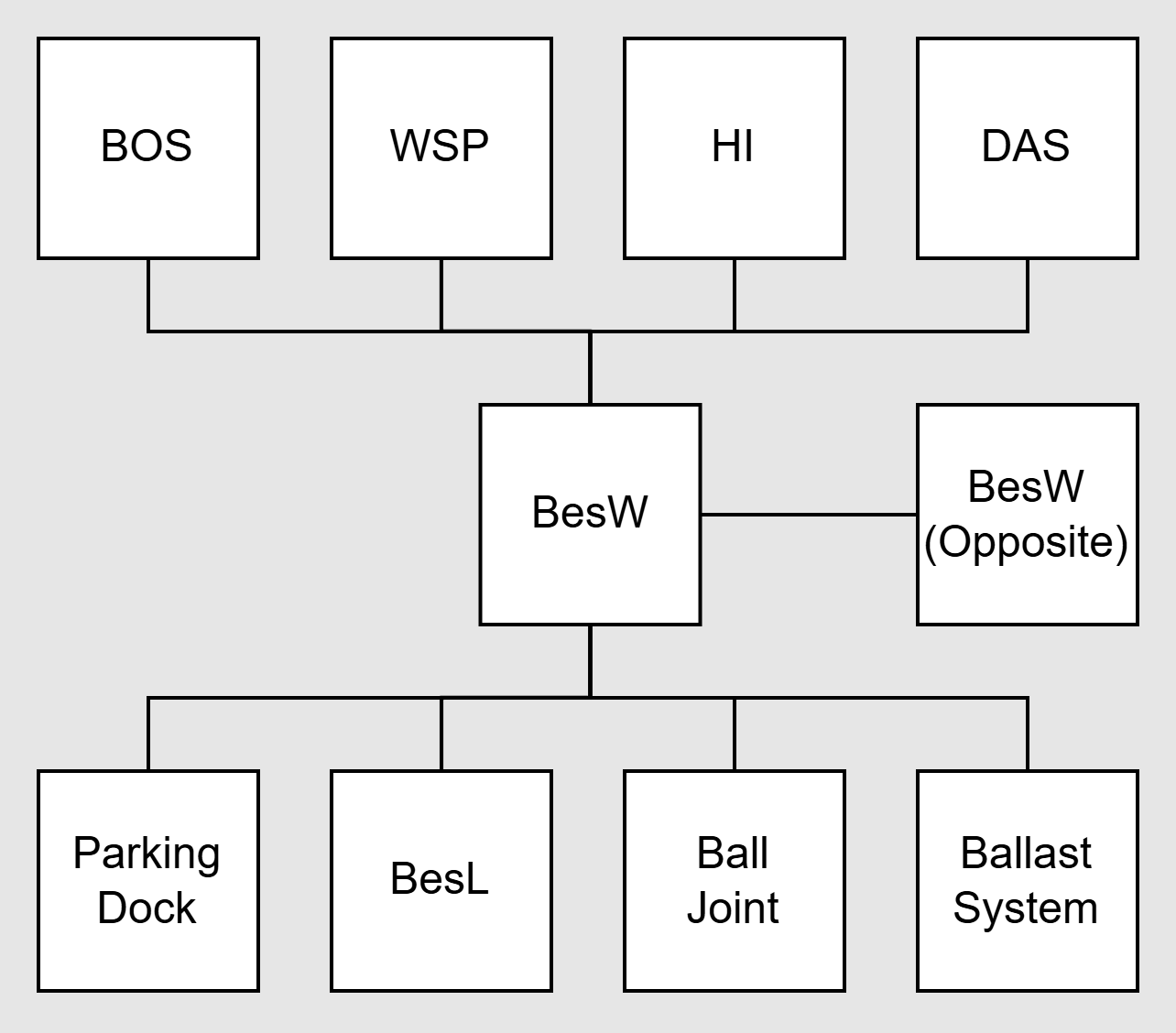}
  \caption{An overview of the control system of the Maeslant Barrier}
  \label{overviewcontrolmaeslant}
\end{figure}

BesW receives commands from BOS (Beslis en Ondersteunend Systeem) and WSP (Waterstanden Paneel). They are responsible for the decisions to close or open the barrier. HI represents the interface of operators manually controlling the system. DAS (Data Acquisition System) can give commands to BesW in order to perform functional tests during maintenance.
BesW North is directly connected to these systems, while BesW South receives the commands from North.
See Fig.~\ref{overviewcontrolmaeslant} for an overview. 

We explain the subsystems of BesW in further detail. All subsystems can, next to BesW, also be controlled by the Motor Control Center (MCC) if BesW allows for it. Through MCC, more local control of the devices can be taken.
Therefore, each process can require slightly different behaviour from each part of the system depending on the source of the instruction, and needs to be modelled separately.

\subsection{Parking Dock}
\begin{figure}[htbp]
  \centering
  \includegraphics[width=0.5\textwidth]{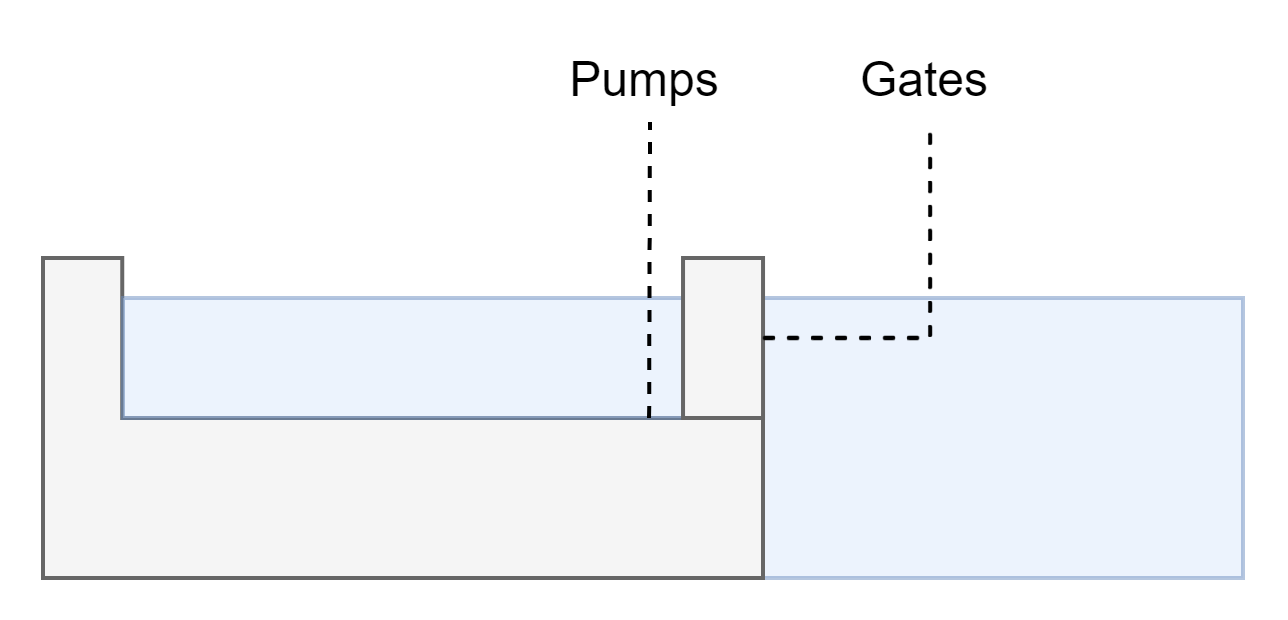}
  \caption{Parking dock components}
  \label{fig:parking-dock}
\end{figure}

Each parking dock has water pumps and a door (see~Fig.~\ref{fig:parking-dock}). 
BesW controls the parking dock devices and takes measurements from it. The most important measurement taken in the parking dock is the difference between the water level in the dock and the river. In the opening procedure, the door only opens once the water levels match. The pumps and gates in the docks's door are used to achieve that.

The door contains catches, winches, and a pawl to open and close the door. It is powered by a power pack. The door also contains limit switches to indicate whether the door is open or closed.

\subsection{BesL}
BesL controls the locomobile responsible for the lateral movement of the retaining wall. BesW will instruct BesL on its movement, its spring settings, and its ventilator. It will also inform BesL if the wall has landed on the river bed. 
BesL in return will inform BesW on its position, its spring setting when enabled, and whether it detects a fire in the locomobile.

\subsection{Ball joint}
BesW can instruct the ball joint to be raised to facilitate maintenance and minimise wear and tear on the barrier. It contains six jacks to raise the ball joint, and ventilators to cool the building containing the ball joint.

\subsection{Ballast system}
\begin{figure}[hbtp]
    \centering
    \includegraphics[width=0.9\linewidth]{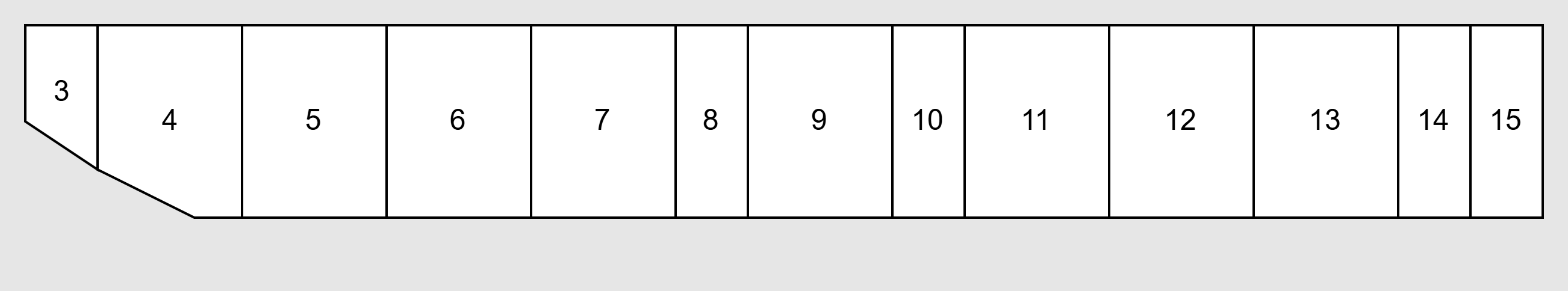}
    \caption{Ballast compartments}
    \label{fig:ballast}
\end{figure}

The retaining wall contains a ballast system to submerge into and emerge from the river. Valves and pumps transfer water in and out of the door compartments (see Fig.~\ref{fig:ballast}). To keep the wall balanced while emerging and submerging, it is vital to coordinate the various pumps to properly distribute the water weight between each compartment. If one or multiple pumps fail (up to some limit), the system should keep working.

Some important measurements the ballast system reports to BesW are the wall slit (the distance between the bottom of the wall and the rived bed), the horizontal tilt of the wall, the water level and volume in each compartment, and the tension exerted by the river on the wall for each side of the wall. More details on the forces on the retaining wall and the data signals of each compartment can be found in \cite{ponsioenInvestigationDevelopmentDigital2023}. 

\section{mCRL2 language}
mCRL2 \cite{bunteMCRL2ToolsetAnalysing2019b} is a language and a set of tools to model behaviour of concurrent and distributed systems, based on process algebra theory and abstract data types. We give a minimal example of the modelling capabilities and how one can formulate properties to be verified on the model below. The purpose of this is just to get an impression of the style in which behaviour and properties are described. The language is described in full in \cite{grooteModelingAnalysisCommunicating2014a}.

\subsection{Specification}

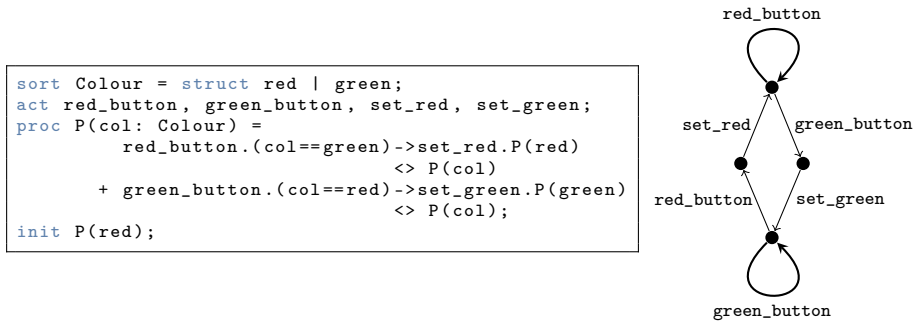
\begin{figure}[h]
\begin{center}\hspace{-0.8cm}
  \begin{minipage}{.67\textwidth}
    \begin{lstlisting}
sort Colour = struct red | green;
act red_button, green_button, set_red, set_green;
proc P(col: Colour) = 
         red_button.(col==green)->set_red.P(red) 
                                <> P(col)
       + green_button.(col==red)->set_green.P(green) 
                                <> P(col);
init P(red);
    \end{lstlisting}
  \end{minipage}
  \hspace{-0.0cm} %space between listing bloc and the figure
  \begin{minipage}{0.2\textwidth}
    \begin{tikzpicture}[
        node distance=2cm,
        state/.style={circle, draw, thick, minimum size=0.5cm},
        arrow/.style={->, >=stealth, thick},
      ]
     \tikzstyle{state} = [circle, fill, inner sep=0pt, minimum size=5pt]
      % Initial state
      \node[state] (s0) {};
      \node[state] (s1) [below of=s0] {};
            \node[state] (s01) [yshift=0.4cm, xshift=-1cm, below right of=s0] {};
            \node[state] (s10) [yshift=0.4cm, xshift=1cm, below left of=s0] {};

    \path[->] 
              (s1) edge  node[xshift=-20pt] {\texttt{\scriptsize red\_button}} (s10)
              (s10) edge  node[xshift=-15pt] {\texttt{\scriptsize set\_red}} (s0)
              (s0) edge  node[xshift=25pt] {\texttt{\scriptsize green\_button}} (s01)
              (s01) edge  node[xshift=20pt] {\texttt{\scriptsize set\_green}} (s1);

     % Create two self-loops, one above and one below
      \path (s0) ++(-1,1) coordinate (helper1);
      \path (s0) ++(1,1) coordinate (helper2);

      \path (s1) ++(-1,-1) coordinate (helper4);
      \path (s1) ++(1,-1) coordinate (helper5);

      % Draw the loops
      \draw[arrow] (s0) .. controls (helper1) and (helper2) ..
      node[above] {\texttt{\scriptsize red\_button}} (s0);

      \draw[arrow] (s1) .. controls (helper4) and (helper5) ..
      node[below] {\texttt{\scriptsize green\_button}} (s1);

    \end{tikzpicture}
  \end{minipage}
  \end{center}
  \caption{A process specification recursively setting traffic lights, and its LTS}
  \label{lst:sending-bits}
\end{figure}

In Fig.~\ref{lst:sending-bits}, a specification describes a traffic light that can be turned red or green. Using the keyword \lstinline|sort| the colour data sort is defined, giving two options \lstinline|red| and \lstinline|green|. With the keyword \lstinline|act| four actions are defined that define the inputs or outputs of the system. In this case, they specify pressing buttons and setting a traffic light to a certain colour.

By the keyword \lstinline|proc| a process \lstinline|P| is defined that waits for a button press to instruct a light to go to a particular colour, and if not showing that colour, sets the traffic light to it, after which it repeats this behaviour keeping the shown colour in the variable \lstinline|col|. The dot represents sequential composition, the plus is alternative composition, or a choice between behaviours, and the if-then-else is denoted by \texttt{\_->\_<>\_}.
mCRL2 supports far more, such as parallel behaviour, a wide range of data types, time, and probabilities.

\subsection{Modal formulae}
Properties of a model can be described using modal formulae in the modal $\mu$-calculus. The mCRL2 toolset can then verify these properties against the behavioural model.
We give some limited examples to get a feel for modal formulae. 

We discern two categories in which most modal formulae fall: safety properties and liveness properties. Safety properties express that nothing bad can ever happen. For instance, the traffic light may never be set to red twice without being set
to green in the meantime. This is expressed by \lstinline{[true*.set_red.!set_green*.set_red]false}. It essentially says that after an arbitrary sequence of actions (\lstinline{true*}) two actions \lstinline{set_red} cannot follow each other with a sequence not containing the action \lstinline{set_green} in between, expressed by \lstinline{(!set_green)*}.

Liveness properties say that something good will eventually happen. One example is that after pressing the red button, the light will be set to red, formalised for instance by \lstinline{[true*.red_button.!set_red*]<true*.set_red>true}. 

The modal $\mu$-calculus allows for explicit minimal and maximal fixed points that, when used in a nested way, even allows fairness properties to be formulated. In the minimal and maximal fixed point operators, parameters can be used which makes
the language the most expressive modal language currently in existence, more than sufficient to formalise all the properties we encountered. 
For instance, it allows for linearly expressing LTL properties \cite{DBLP:journals/tcs/CranenGR11}.

\section{Model BesW}
\begin{figure}
\begin{lstlisting}
proc Process_HI_OpenDoor(processMode: ProcessMode, system: MainSystem) =
        (processMode == active) -> (
            (system == dock) -> (
                Dock_Door_ChooseWinchOpen_InOrder(requireCatchTimer=false).
                Dock_Door_Open(requireDevicesReady=true).
                Dock_Door_CheckOpened
            ) +
            (system == besl) -> (
            skip
            ) +
            (system == joint) -> (
                Joint_Jack_Control
            ) +
            (system == ballast) -> (
                Ballast_Comps_DrainWithPumps_General.
                Ballast_Comps_DrainWithRestPumps
            )
        ) +
        (processMode == finished) -> (
            (system == dock) -> (
                skip
            ) +
            (system == besl) -> (
                skip
            ) +
            (system == joint) -> (
                Joint_Jack_Control
            ) +
            (system == ballast) -> (
                Ballast_Comps_DrainWithPumps_General.
                Ballast_Comps_DrainWithRestPumps
            )
        ) +
        (processMode == stopped) -> (
           (system == dock) -> (
                Dock_Door_OpenStop
            ) +
            (system == besl) -> (
                skip
            ) +
            (system == joint) -> (
                Joint_Jack_Control
            ) +
            (system == ballast) -> (
                Ballast_Comps_DrainWithPumps_General.
                Ballast_Comps_DrainWithRestPumps
            )
        );
\end{lstlisting}
\caption{A fragment of mCRL2 specification of BesW}
\label{tab:BESW_mCRL2}
\end{figure}
The specification of the main control system of the Maeslant Barrier is described internally at Rijkswaterstaat in the Detail Software Ontwerp (DSO) document~\cite{rijkswaterstaatDetailSoftwareOntwerp2021}. The document contains a very detailed description of the specified behaviour of BesW in the form of 15.000 lines of low-level PLC code. In addition to that, several control tables describe the pump and valve settings of each compartment in the retaining wall, depending on the water levels, the tilt of the wall, and the phase of closure. The DSO is viewed within Rijkswaterstaat as the authoritative document, although it is quite inaccessible and has never been formally checked. The modelling in mCRL2 revealed that the DSO contains at least one issue where, when pumps malfunction, more pumps can be on than the electricity budget allows. 

Our model is a translation of the DSO and the control tables into an mCRL2 specification. The mCRL2 description is around 5.000 lines, including empty lines. The size of the state space is $4.98*10^{14}$ as calculated using \verb|lpsreach|.

Besides mended system behaviour, the mCRL2 model deviates from the DSO in the following aspects:
\begin{itemize}
    \item We separated control instructions from `coupled' control into two new levels: one for instructions from BOS and one for instructions from human interfaces.
    \item Actions corresponding to notifications to human operators are omitted to achieve a smaller state space and focus on the control flow itself.
    \item Sensor measurements are abstracted to single relevant values. For instance, water level sensor data and ballast system data often follow some complex equations based on many sensors, so we only consider the values that might result from such a computation.
    \item The control of devices is modelled with single actions, instead of using precise control, where, for instance, an output voltage must be high for a certain amount of time for which the DSO uses explicit timers. 
    \item Measurements such as water levels are also discretised to aid in verification.
    \item The data protocol used for communication with BOS is not modelled.
\end{itemize}

The model follows the PLC structure where every major process can be in one of the following modes: active, stopped, or finished (see Figures \ref{fig:operational-processes} and~\ref{overviewcontrolmaeslant}). It is important to explain the difference between `processes' in the DSO and `processes' in mCRL2. Processes in the DSO are procedures that indicate the current state to a distributed system so that each system can decide what actions it can and should take.
Meanwhile, processes in mCRL2 are understood to describe the systems, and they decide themselves how each system should behave.

BOS only allows one active process at any time (with some small exceptions) and only allows logical sequences of processes. For example, the retaining wall cannot be instructed to move out, when the door of the dry dock is not open, contrary to what DSO allows. This is done to not let the state space grow too much. Rare but possible sequences are allowed. It is allowed to start closing the door while it is not yet fully opened. 

In order to give an idea of how the specification looks like, we give a small fragment in Fig. \ref{tab:BESW_mCRL2} which is exemplary for the whole specification. Influenced by the structure of the DSO -- and one can discuss whether this is the most optimal -- the specification repeats itself in cycles, where in each cycle each system performs its tasks. The currently active system is indicated by the process parameter \lstinline{system}. The shown process opens the door of the dry dock started from HI. For instance, when the process is in the active mode, the \lstinline{dock} executes three subprocesses to choose the winches, open the door and check the door successfully opened, indicated by processes described elsewhere in the specification. In the same mode, the \lstinline{joint} system should run the \verb|Joint_Jack_Control| process also described at another spot in the specification. The \lstinline{besl} system does not perform any task, indicated
by the explicit action \lstinline{skip}. All processes have access to a set of global data variables maintained in a parallel thread. 

\section{Checking behavioural properties}
To guarantee that formal specifications are of the required quality, it is necessary to verify behavioural properties \cite{DBLP:journals/scp/BrandG15,DBLP:journals/sttt/OsaiweranSHGR16}. 
We checked 40 properties. These properties were inferred from the DSO, our understanding of the desired behaviour, and from conversations with top experts on storm surge barriers and the top experts on the Maeslant Barrier.
We provide a number of examples of these properties. Most of them turned out to be valid immediately, 
some that required some improvements in the model or the requirement, and one turned out to be actually fundamentally incorrect. 
All formal properties are public and can be found in~\cite{beersSpecificationVerificationMain2024}.

% Ik snap de bedoeling van onderstaande niet. 
%The properties operate under the following processes described in Fig.~\ref{table:processes}. The category `Parking Dock Devices' is used when devices are tested across various processes\todo{Ik snap hier geen hout van}.

%\def\arraystretch{1.1}
%\setlength\tabcolsep{10pt}
%\begin{table}[htbp]
%    \centering
%    \begin{tabular}{|l|l|}
%    \hline
%      Reach Rest Level& Reach the Rest level of water in the dock. \\
%      Equalise Level& Equalise river water level with the level of water in the dock.\\
%      Close Door& Close the dock door.\\
%      Open Door& Open the dock door.\\
%      Move In& Move the retaining wall laterally into the dock.\\
%      Move Out& Move the retaining wall laterally out of the dock.\\
%      Emerge& Move the retaining wall vertically up out of the river.\\
%      Submerge& Move the retaining wall vertically down into the river.\\
%      \hline
%    \end{tabular}
%    \caption{Operational processes}
%    \label{table:processes}
%\end{table}

\subsection{Properties that were immediately valid}

We give some examples of the 36 formulas that were immediately valid. We also give the modal formulas, 
and give a short explanation of the encoding the modal $\mu$-calculus.  

\subsubsection{Open Door: Finishing condition.} The process Open Door will finish if and only if at least two limit switches indicate the door is opened. The formula says that whenever the process Open Door is started and not ended, represented
by \lstinline{(!internal_controlEnd)*}, and a list of lists of boolean \lstinline{doorOpenedSensors} is received from the
dock door, then the \lstinline{processMode} is equal to \lstinline{finished} exactly if at least two different booleans in \lstinline{doorOpenedSensors} are set to true, indicating that at least two switches indicate that the door is open. 

\begin{lstlisting}
[ true*.
  internal_controlStart(operational, processOpenDoor, active).
  (!internal_controlEnd)*
]
(forall doorOpenedSensors: List(List(Bool)). val(#doorOpenedSensors == 3 && 
     (forall i: Nat. i < 3 => #(doorOpenedSensors.i) == 2)) =>
  [
    input_dockDoorOpened(doorOpenedSensors) .
    (!internal_controlEnd)* .
    internal_controlEnd .
    (!internal_controlEnd)*
  ]
  (forall processMode: ProcessMode.
    [internal_controlStart(operational, processOpenDoor, processMode)]
    val(
      (processMode == finished)
      ==
      (exists i,j,i',j': Nat. (i < 3 && j < 2 && i' < 3 && j' < 2 && i != i' && doorOpenedSensors.i.j && doorOpenedSensors.i'.j')
) ) )
\end{lstlisting}

\subsubsection{Move Out: Trimming.} During the active mode of process Move Out, trimming will not be active.
The encoding is straightforward and says that when the process Move Out is started, and not finished, the action
indicating that triming is active cannot happen. 
\begin{lstlisting}
[
  true* .
  internal_controlStart(operational, processMoveOut, active) .
  (!internal_controlEnd)* .
  internal_trimmingActive
]
false
\end{lstlisting}

\subsubsection{Submerge: Spring setting to K2.} During the active mode of process Submerge, the spring setting of BesL will be changed to K2 when the wall slit is below 3.5m. The encoding says that when the process Submerge is active, and when it
sets a spring setting, then this spring setting is K2 when the wall slit was measured to be smaller than 3.5m, which
slightly diverges from the informal requirement, but is defensible as the spring setting is changed anyhow.

\begin{lstlisting}
[
  true* .
  internal_controlStart(operational, processSubmerge, active) .
  (!internal_controlEnd)*
]
(forall wallSlitIndex: Nat. val(wallSlitIndex < #wallSlitList) =>
  [
    input_wallSlit(wallSlitList.wallSlitIndex) .
    (!internal_controlEnd)*
  ]
  (forall springSetting: SpringSetting.
    [output_beslSpringSetting(springSetting)]
    val(wallSlitList.wallSlitIndex < 350/100 => (springSetting == K2))
) )
\end{lstlisting}

\subsection{Properties that became valid after refinement}
Here we report on some properties we were asked to check but which had to be rephrased to become valid.

\subsubsection{Reach Rest Level, Active: Opening gates.}
The first property says that the gates will be opened when the dock level is under -2.4 m. After analysing counter examples, we refined the requirement to the following: During the active mode of process Reach Rest Level in the Operational phase, the gates will be opened if the dock level is under -2.4 m, the ball joint is lowered, and the gates are not opened yet, formulated
in the modal formula below. 

\begin{lstlisting}
[ true* . 
  internal_controlStart(operational, processReachRestLevel, active) . 
  (!internal_controlEnd)*
]
(forall gatesOpened: List(Bool). val(#gatesOpened == 4) =>
  [
    input_dockGatesOpened(gatesOpened) . 
    (!internal_controlEnd)*
  ]
  (forall dockLevelIndex: Nat. val(dockLevelIndex < #dockLevelList) => 
    [
      input_dockLevel(dockLevelList.dockLevelIndex) . 
      (!internal_controlEnd)*
    ]
    (forall jackZero: Bool.
      [
        input_jointJackZero(jackZero) . 
        (!internal_controlEnd)*
      ]
      (forall gatesOpen: List(Bool). val(#gatesOpen == 4) => 
        [output_dockGatesOpen(gatesOpen)]
        val(dockLevelList.dockLevelIndex <= -240/100 => (forall g: Nat. g < 4 
            => (jackZero && !(gatesOpened.g) => gatesOpen.g)
) ) ) ) )

\end{lstlisting}

\subsubsection{Equalise Level, Active: Finishing condition.}
We were asked to check the property that the process Equalise will finish, if at least two gates are opened.
The property had to be rephrased to: The process Equalise Level will finish if and only if the timer for equalising has expired or the dock level is equalised with the river, as formalised by the following formula. 

\begin{lstlisting}
[
  true* . 
  internal_controlStart(operational, processEqualiseLevel, active) . 
  (!internal_controlEnd)*
]
(forall riverLevelIndex: Nat. val(riverLevelIndex < #riverLevelList) => 
  [
    internal_riverLevel(riverLevelList.riverLevelIndex) . 
    (!internal_controlEnd)*
  ]
  (forall dockLevelIndex: Nat. val(dockLevelIndex < #dockLevelList) => 
    [
      input_dockLevel(dockLevelList.dockLevelIndex) . 
      (!internal_controlEnd)*
    ]
    (forall stateTimerEqualiseLevel: TimerState.
      [
        state''(timer_equalise, stateTimerEqualiseLevel) . 
        (!internal_controlEnd)* .
        internal_controlEnd .
        (!internal_controlEnd)*
      ]
      (forall processMode: ProcessMode. 
        [internal_controlStart(operational, processEqualiseLevel, processMode)] 
        val(
          (processMode == finished)
          ==
          (stateTimerEqualiseLevel == expired || dockLevelEqualised(dockLevelList.dockLevelIndex, riverLevelList.riverLevelIndex))
) ) ) ) )
\end{lstlisting}

\subsubsection{Parking Dock Devices: Catch.}
This property stated that during processes where the dock door is closed, the catch will be fastened when positive head, i.e., the water level in the dock, is too low. Also, the catch will be loosened when positive head in the dock is sufficiently high.

This property was purposefully described to be false by subject matter experts to check if formal verification could detect it. The condition that was missing was that fastening the catch also requires that the limit switches indicate the door is closed. This is formalised below. 

\begin{lstlisting}
[true*]
(forall process: Process, processMode: ProcessMode.
  [
    internal_controlStart(operational, process, processMode) . 
    (!internal_controlEnd)*
  ]
  (forall riverLevelIndex: Nat. val(riverLevelIndex < #riverLevelList) => 
    [
      internal_riverLevel(riverLevelList.riverLevelIndex) . 
      (!internal_controlEnd)*
    ]
    (forall dc: List(Bool). val(#dc == 2) =>
      [
        input_dockDoorClosed(dc) . 
        (!internal_controlEnd)*
      ]
      (forall dockLevelIndex: Nat. val(dockLevelIndex < #dockLevelList) => 
        [
          input_dockLevel(dockLevelList.dockLevelIndex) . 
          (!internal_controlEnd)*
        ]
        (forall catchFasten: Bool.
          [output_dockCatchFasten(catchFasten)]
          val(
            (
              process == processEqualiseLevel || 
              process == processReachRestLevel ||
              (process == processCloseDoor && processMode == finished)
            )
            =>
            (
              (dc.0 || dc.1)
              &&
              (riverLevelList.riverLevelIndex - dockLevelList.dockLevelIndex < 70/100)
              =>
              catchFasten
            )
            &&
            (
              (riverLevelList.riverLevelIndex - dockLevelList.dockLevelIndex > 100/100) 
              => 
              !catchFasten
) ) ) ) ) ) )
\end{lstlisting}

\subsection{An Invalid Property}
We were asked to check one property that we were unable to reformulate into a valid variant. 
The property states that during the Operational phase, if BesL indicates that the retaining wall is not well positioned, then the pumps of the parking dock will not be enabled. It is encoded below. 
After some discussion, it turned out that this functionality was deprecated and part of an earlier specification document. 

\begin{lstlisting}
[true*]
(forall process: Process, processMode: ProcessMode.
  [
    internal_controlStart(operational, process, processMode) .
    (!internal_controlEnd)* .
    input_beslFinePositioned(false) .
    (!internal_controlEnd)*
  ]
  (forall p: Nat. val(p < 3) =>
    [output_dockPumpEnable(p, true)] false
  )
)
\end{lstlisting}
\subsection{Tools}
We verified the properties against the model by first transforming the mCRL2 specification into a Linear Process Specification (LPS). Tools are then applied to reduce the LPS. 
Next, the LPS is combined with a modal formula into a Parameterized Boolean Equation System (PBES). After applying tools to reduce the PBES, we solve it. The following commands were used where the specification is given as \lstinline{besw.mcrl2}
and the property as \lstinline{property.mcf}.
\begin{lstlisting}[language=bash]
mcrl22lps -vn -Q0 besw.mcrl2 | lpsconstelm | lpssumelm | lpsrewr | lpsparelm | lpsstategraph - spec.lps
lps2pbes -m -f property.mcf spec.lps | pbesconstelm | pbesparelm | pbesstategraph -Q0 | pbessolve -Q0 -rjittyc --threads=4
\end{lstlisting}

Verification was performed on a system belonging to Rijkswaterstaat. mCRL2 version 202407.1 was used.
The system has two AMD EPYC 9254 24-Core processors.
Each CPU has 12 DDR5 DIMMs of 64GB for a total of 1.5TB of memory. We were able to verify each property within forty minutes. 

\section{Conclusions}
We have shown that we can fully model BesW and that verification of various behavioural properties is more than doable. Although detailed and precise, the specification is of limited length and can therefore neatly serve as the specification of any BesW software controller to be made. 
This provides an overwhelming positive answer to the primary research question whether such systems can be modelled and verified.
Without any hesitation, we can state that in the hands of qualified specifiers, guided by domain experts, formal methods are more than capable to precisely
model the software of systems of the complexity of storm surge barriers and prove that these specifications are of quality. It is important to find the right level of modelling to only grasp key behaviour~\cite{DBLP:journals/stvr/GrooteKO15}. Here, this means primarily avoiding parallel behaviour, abstractly describing the essence of the interactions of BesW with the outside world, and avoiding specifying log messages. 

We based the specification on the DSO. This was truly helpful, as -- despite its relative inaccessible nature -- it provided many behavioural details that we otherwise only could have guessed. This also had a disadvantage in that our mCRL2 specification still resembles the DSO by using a super-loop as most PLC systems do. Removing the loop could make the specification even more compact and accessible. Also, the language mCRL2 would benefit from some adaptations to improve the conciseness and readability of the specification. 

The biggest hurdle left for further adoption of formal verification is education. The concepts of modelling and verification are still relatively unknown to storm surge barrier experts and software engineers programming the storm surge barriers. Learning to create models and formal requirements requires practice. But, as the advantages of verified software specifications are so immense, this issue will be overcome.

\bibliographystyle{splncs04.bst}
\bibliography{bibl.bib}

\end{document}